\documentclass{article}
\usepackage{amsmath,graphicx,mlspconf}

\usepackage{cite}
\usepackage{amssymb,amsfonts,amsthm}
\usepackage{algorithmic}
\usepackage{textcomp}
\usepackage{xcolor}
\usepackage{mathtools}
\usepackage{bbm}
\usepackage{lipsum} 
\usepackage{wrapfig}
\usepackage{caption}
\usepackage{multirow}
\usepackage{xargs}
\usepackage{url}
\usepackage{booktabs}
\usepackage{tabularx}
\usepackage{nicefrac}
\usepackage[ruled,vlined]{algorithm2e}
\def\BibTeX{{\rm B\kern-.05em{\sc i\kern-.025em b}\kern-.08em
    T\kern-.1667em\lower.7ex\hbox{E}\kern-.125emX}}

\newcommand{\gray}[1]{\textcolor{gray}{#1}}

\newenvironment{sketch}{%
  \renewcommand{\proofname}{Proof Sketch}\proof}{\endproof}
\newtheorem{proposition}{Proposition}
\newtheorem{remark}{Remark}

\definecolor{lb}{RGB}{31,119,180}
\definecolor{forestgreen}{HTML}{008000}

\usepackage{amsmath,amsfonts, amssymb, bm, mathtools}

\def\subsecref#1{subsection~\ref{#1}}
\def\propref#1{Proposition~\ref{#1}}

\def\figref#1{Figure~\ref{#1}}

\def\eqref#1{Equation~(\ref{#1})}
\def\tabref#1{Table~\ref{#1}}

\def\1{\bm{1}}

\def\vmu{{\bm{\mu}}}

\def\vx{{\bm{x}}}
\def\vy{{\bm{y}}}

\DeclareMathAlphabet{\mathsfit}{\encodingdefault}{\sfdefault}{m}{sl}
\SetMathAlphabet{\mathsfit}{bold}{\encodingdefault}{\sfdefault}{bx}{n}

\DeclareMathOperator*{\argmax}{argmax}

\DeclareMathOperator*{\simplies}{\DOTSB\Longrightarrow}

\copyrightnotice{Author copy of accepted manuscript}

\toappear{2023 IEEE International Workshop on Machine Learning for Signal Processing, Sept.\ 17--20, 2023, Rome, Italy}

\title{Greedy online change point detection}%
\name{Jou-Hui Ho,  Felipe Tobar\thanks{We thank Google and the following ANID-Chile grants: Fondecyt-Regular 1210606, Advanced Center for Electrical and Electronic Engineering (Basal FB0008) and Center for Mathematical Modeling (Basal FB210005)}}
\address{Universidad de Chile}

\begin{document}
\ninept

\maketitle

\begin{abstract}
    Standard online change point detection (CPD) methods tend to have large false discovery rates as their detections are sensitive to outliers.
    To overcome this drawback, we propose \textit{Greedy Online Change Point Detection} (GOCPD), a computationally appealing method which finds change points by maximizing the probability of the data coming from the (temporal) concatenation of two independent models. 
    We show that, for time series with a single change point,  this objective is unimodal and thus CPD can be accelerated via \textit{ternary search} with logarithmic complexity. 
    We demonstrate the effectiveness of GOCPD on synthetic data and validate our findings on real-world univariate and multivariate settings.
    \end{abstract}

\begin{keywords}
change point detection, Gaussian process
\end{keywords}

    \section{Introduction}
    The need to detect abrupt changes in time series is relevant in many fields, such as genomics \cite{genomic}, speech recognition \cite{speech}, finance \cite{finance2} and medical signals monitoring \cite{epileptic, heart_rate, eeg_cpd2021}. 
    Existing CPD methods can be categorized into Bayesian or non-Bayesian. 
    The former \cite{saaturras2010, agudelo-espana20a, alami20a, BOCPDMS} build on Bayesian Online Change Point Detection (BOCPD) \cite{adams2007bayesian}, that models the change-point probability recursively while providing uncertainty quantification for each detection.
    BOCPD's change point probability depends on the probability of the streaming observations, therefore, noisy data may lead to a large false positive rate.
    %
    %
    Non-Bayesian methods mainly rely on a likelihood ratio  \cite{481608, Gustafsson2000AdaptiveFA}, which also leads to false discoveries under the presence of outliers.

    Our aim is to develop an online CPD algorithm in the \textit{unsupervised} learning context, i.e., there are no labeled change points available during training. An intuitive way toward this aim is to analyze the arriving samples and decide whether the so-far observed data are more likely to follow a single model or two different models operating on consecutive, non-overlapping temporal segments. Inspired by common sense, we argue that when the observations appear to be distributed differently from the previous ones, the detector should not notify the change instantly. Instead, it should wait for more observations and, if the new regime is sustained, then notify a change as an informed decision.
    
    We propose \textit{Greedy Online Change Point Detection} (GOCPD) a method suited for real-world implementation
     since it does not require prior knowledge of the number of change points or states unlike previous works \cite{Fearnhead&Liu, saaturras2010}. Our contribution is threefold: $(i)$ the formulation of the online CPD through two simple likelihood-based criteria to search and declare a change point; $(ii)$ a computationally-efficient approach to search the candidate change point with ternary search, which operates in logarithmic time; and $(iii)$ experimental validation of the ability of GOCPD to  detect changes in both synthetic and real-world data (uni- and multi-variate).

\section{Background} \label{sec:background}

\subsection{Notation and problem statement} \label{subsec:problem}

    Let $\mathcal{D} = \{(x_t, y_t)\}_{t=1}^T$ denote observations of a time series, where $x_t \in \mathbb{R}^{D}$ and $y_t \in \mathbb{R}^{C}$ are the input and output at time  $t$ respectively, and  $C$ is the number of channels. %
    Also, let $\mathcal{D}_{t_i:t_j} = \{(x_t, y_t)\}_{t=t_i}^{t_j}$ denote the observations between $t=t_i$ and $t=t_j$. 
    Consider that the distribution of this time series changes abruptly at timestamps $\{c^*_1, c^*_2, \cdots, c^*_n\}$, so that $ \forall i \in \{2, \dots, n-1\}$, the sets $ \mathcal{D}_{c^*_{i-1}:c^*_{i}-1}$ and $\mathcal{D}_{c^*_{i}:c^*_{i+1}-1}$ are independent and generated by different models.
    %
    We assume $c^*_1 = 0$ under the realistic assumption that the first segment starts at the first observation.
    We denote the detected changes as $\tilde{\mathcal{C}} = \{ \tilde{c}_1, \cdots, \tilde{c}_m \}$, where $n=m$ does not necessarily hold.
    At each $t$, we denote $\tilde{c} : \mathbb{R} \to \mathbb{R}$ the function that maps each $t$ to the last detected change point before time $t$, e.g., for $t\in(\tilde{c}_1, \tilde{c}_2)$, it holds that $\tilde{c}(t) = \tilde{c}_1$, which is not necessarily equal to $c^*_1$.
    %
    We consider that the models before and after the change point are unknown.

\subsection{Previous work} \label{subsec:previous_Work}
    
    \textbf{Bayesian methods} provide uncertainty measures for each detection. An example is the Bayesian Online Change Point Detection (BOCPD)  \cite{adams2007bayesian, Fearnhead&Liu}, which is based on a recursive formulation of the \emph{run-length}, i.e., the elapsed time since the last change. BOCPD monitors the \emph{run-length} $r_t$ at time $t$, that is, the time since the last change point. This requires to define the conditional prior $p(r_t|r_{t-1})$ as well as the change and grow probabilities, respectively $p(r_t=0)$ and $p(r_t=r_{t-1} +1 )$. In real-world settings, these quantities are often unknown, context dependent and difficult to parameterize. BOCPD variants  incorporate Gaussian processes \cite{saaturras2010} or latent variables \cite{morenomunoz2019continual}, to increase robustness  \cite{BOCPDMS, knoblauch2018doubly} or to predict change points \cite{agudelo-espana20a}.

    \textbf{Non-Bayesian methods} compare two consecutive time intervals of fixed window size \cite{cusum1, cusum2, SPLL}, and apply a Generalized Likelihood Ratio (GLR), an extension of the CUSUM strategy \cite{cusum1}.
    Recently, \cite{newma} introduced NEWMA, a model-free method that summarizes old and new data in two statistics using the Exponential Weighted Moving Average algorithm with different forgetting factors. NEWMA declares a change based on the Maximum Mean Discrepancy.
    
    For a survey on CPD methods, we refer to \cite{survey, evaluation}.

\section{Proposed model} \label{model}

    Our proposal, termed Greedy Online Change Point Detection (GOCPD), aims to overcome the drawback of previous CPD methods, which mostly depend on the probability of the last observed data and thus are prone to large false discovery rates. The key component of GOCPD is the use of Gaussian models to search for the timestamp that splits the so-far observed data into the two most-likely segments. We call this timestamp the \emph{candidate change point}. Then, GOCPD evaluates if this split provides enough discrepancy with respect to the data coming from one single model for a sufficiently long period to approve the candidate as a detection.

\subsection{Locating the the candidate change point}    
    
    Consider datapoints $\mathcal{D}_{\tilde{c}(t):t} \subseteq \mathcal{D}$ with a change point at time $c^* \in \{\tilde{c}(t) + 1, \tilde{c}(t) + 2, ..., t-1\}$. An intuitive procedure for online  CPD is to analyze the observations available up to time $t$, and identify the change point at a timestamp $\tau$ \textit{previous} to the current time $t$, so that the collections $\mathcal{D}_{\tilde{c}(t):\tau}$ and $\mathcal{D}_{\tau:t-1}$ are independent and given by different models. 
    
    We will assume that the probability of the data coming from independent models (i.e., $\mathcal{D}_{\tilde{c}(t):\tau-1}$ comes from a pre-change model $m_1$ and $\mathcal{D}_{\tau:t}$ comes from a post-change model $m_2$), is larger than that of the data coming from a single model, denoted as $m_0$. 
    For numerical stability, we consider the log-likelihood ratio given by $L_t(\tau) = \log p(\mathcal{D}_{\tilde{c}(t):\tau-1}|\hat\theta_1) + \log p(\mathcal{D}_{\tau:t}|\hat\theta_2) - \log p(\mathcal{D}_{\tilde{c}(t):t}|\hat\theta_0)$, where $\hat\theta_1, \hat\theta_1$ and $\hat\theta_2$ are the maximum likelihood estimators of the parameters in $m_0, m_1, m_2$, respectively.
    
    Remarkably, although $L_t(\tau)$ is a log-likelihood ratio, since the last term $\log \Bar{p}(\mathcal{D}_{\tilde{c}(t):t}|\hat\theta_0)$ is constant for a given $t$, the optimization is essentially computed only through the first two terms: $ L_t(\tau) = \log p(\mathcal{D}_{\tilde{c}(t):\tau-1}|\hat\theta_1^{\tau}) + \log p(\mathcal{D}_{\tau:t}|\hat\theta_2^{\tau})$.
    Therefore, this criterion differs from the thresholding of likelihood-ratio-based methods \cite{cusum1, SPLL, cusum2}, since we directly optimize the probability of the data belonging to two independent models instead of a likelihood ratio.
    
    In order to correct for the different amount of observations for each likelihood in $L_t(\tau)$, we consider the average log-likelihood instead, i.e., $\log \Bar{p}(\mathcal{D}|\cdot) = |\mathcal{D}|^{-1}\log p(\mathcal{D}|\cdot)$.
    Thus, we define a \emph{change point location metric} as:
    \begin{equation}
        s_t(\tau) = \log \Bar{p}(\mathcal{D}_{\tilde{c}(t):\tau-1}|\hat\theta_1^{\tau}) + \log \Bar{p}(\mathcal{D}_{\tau:t}|\hat\theta_2^{\tau}).
    \label{eq:criteria}
    \end{equation}

    Therefore, finding the optimal change point location given current observations is given by the maximization of  \eqref{eq:criteria}. We denote its solution as $c_t$:
    \begin{equation}
        c_t = \argmax_{\tau\in \{\tilde{c}(t)+1, \dots, t-1\}} s_t(\tau).
    \label{eq:argmax_criteria}
    \end{equation}

    \subsection{Efficient greedy search of the optimal change point}
    
    A naive way to find $c_t$ is to compute $s_t(\tau)$ over all values of $\tau \in \{\tilde{c}(t), \dots, t\}$ for every $t$, which would take $\mathcal{O}(t)$ for each batch of data.
    Moreover, for small batches  (even for a single datapoint) the optimizer of $s_t$ and $s_{t+1}$ are close to each other, then, computing all values is redundant. 
    To circumvent the time complexity of solving \eqref{eq:argmax_criteria} at each iteration, we rely on the following property of  
 $s_t(\tau)$ in \propref{prop:unimodal}.
    
    \begin{proposition}
    Let $\mathcal{D}_{\tilde{c}(t):t}$ be observations of a time series with a change point at $c^* \in \{\tilde{c}(t) + 1, \tilde{c}(t) + 2, ..., t-1\}$.
    Then, there exists $0 < \delta_1 < c^*$ and $0 < \delta_2 < t - c^*$ such that $\log \Bar{p}(\mathcal{D}_{\delta_1:\tau-1}|\hat\theta_1^{\tau}) + \log \Bar{p}(\mathcal{D}_{\tau:t-\delta_2}|\hat\theta_2^{\tau})$ has a unique maximum at $\tau=c^*$ and monotonically decreases as $\tau$ moves away from $c^*$. \label{prop:unimodal}
    \end{proposition}
    
    \begin{sketch} 
    Since each log-likelihood term in \eqref{eq:criteria} is normalized, for $\tau<c^*$, $\Bar{p}(\mathcal{D}_{\tilde{c}(t):\tau}|\hat\theta_1)$ is approximately constant, and after $c^*$, $m_1$ starts observing data that are not from the original distribution. Thus $\Bar{p}(\mathcal{D}_{\tilde{c}(t):\tau}|\hat\theta_1)$ decreases as $\tau$ moves away from $c^*$. The analysis is analogous for $m_2$. Therefore, the sum of both log-likelihoods has a maximum at $c^*$.
    %
    \end{sketch} 
    
\begin{figure}[t]
        \centering
        \includegraphics[width=0.99\linewidth]{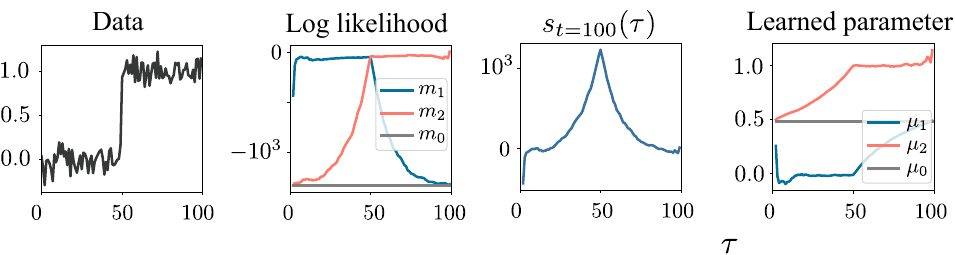}
        \caption{{Unimodal property of $s_t$ under a change of mean (Gaussian).} From left to right: ($i$) observations, ($ii$) log likelihood of $m_{0, 1, 2}$, ($iii$) change point location metric in \eqref{eq:criteria}, and ($iv$)  learned means of $m_{0, 1, 2}$ for each $\tau$.}
        \label{fig:ll_analysis}
    \end{figure}

    \textbf{Example (unimodal behavior of $s_t(\tau)$, Fig.~\ref{fig:ll_analysis}).} We generated a synthetic dataset up to $t=100$ with a change point at $c^*=50$, where $y_{0:49} \sim \mathcal{N}(0, 0.1^2)$ and $y_{50:100} \sim \mathcal{N}(1, 0.1^2)$. 
    This example simulates a batch of streaming data, thus we set $t=100$ for each $\tau$, i.e., the criterion for all $\tau$ is evaluated over the data up to $t=100$.
    We consider a family of models $m_{0,1,2} \in \mathcal{M} = \{ \mathcal{N}(\mu, 0.001^2) \}_\mu$, where $\mu$ is the parameter of the model, which is learned via maximum likelihood.
    Since $\tau=50$ is the true change point, around this value $m_1$ and $m_2$ are trained with their truly generated data, which shows the peak  at $\tau=50$.
    
    Solving \eqref{eq:argmax_criteria} implies finding the maximum of a unimodal function, which is a well-solved problem in the literature.
    %
    %
    Then, instead of evaluating $s_t(\tau)$ on each timestamp, we use \textit{ternary search} \cite{ternary}, a divide-and-conquer algorithm that finds the maximum of a unimodal sequence of size $t$ in $\mathcal{O}(\log(t))$. Therefore, now the per-iteration complexity of the change searching is $\mathcal{O}(\log(t))$. 
    
    However, even using an ternary search, this procedure needs to be repeated for similar data in each iteration of the online scenario. 
    To avoid a redundant search, we build on the following property.
    
    \begin{proposition}
    Given an arbitrary timestamp $t$, let $c_t = \argmax_{\tau} s_t(\tau)$ and $c_{t+1} = \argmax_{\tau} s_{t+1}(\tau)$ be the solutions of \eqref{eq:argmax_criteria} at time $t$ and $t+1$ respectively. Then, it holds that $c_t \leq c_{t+1}$. \label{prop:save_cand}
    \end{proposition} 
    
    \begin{sketch} 
    Given Proposition \ref{prop:unimodal}, $s_t(\tau)$ monotonically increases before the real change point $c^*$, then its maximum is located at or after $c_t$. 
    \end{sketch} 
    
    This property guarantees that, for a resulting candidate change point $c_t$ of an iteration, the optimal change point  the  is never located before $c_t$. Moreover, $c_{t'} \geq c_t, \forall t'>t$.
    Therefore, timestamps before the optimal change location of the previous iteration do not need to be evaluated in the current iteration.
    Following this result, we save the located change point with the highest $s_t$ at each iteration as a \textit{candidate} change point. 
	For the next iteration, GOCPD only evaluates the timestamps after $c_t$ to find the maximum of $s_{t+1}(\tau)$. 
	We call this range of timestamps as the \emph{effective} interval.

\subsection{Accepting the candidate change point} 
    
    Recall that for Gaussian models, the realizations $\mathcal{D}_{\tilde{c}(t):t}$ have a log-probability given by a  \emph{Mahalanobis distance} between the posterior mean and the observations, defined by
    \begin{equation}
        d_m(\mathcal{D}) = \sqrt{(\vy-\vmu_{\vx})^T\Sigma_{\vx\vx}^{-1}(\vy-\vmu_{\vx})}, \label{eq:maha}
    \end{equation} 
    where $\vx = \{x_t\}_t$ and $\vy=\{y_t\}_t$ are the input and output respectively, and $\vmu_{\vx}$ and $\Sigma_{\vx\vx}$ are the posterior mean and variance of the model $m$ on $\vx$. 
    
    Additionally, since  $d_m(\mathcal{D}_{\tilde{c}(t):t})$ depends on the dimension of $\mathcal{D}_{\tilde{c}(t):t}$ ($|\mathcal{D}_{\tilde{c}(t):t}| = t-\tilde{c}(t)+1$, which varies over time) we consider a modified Mahalanobis distance given by $\Bar{d}_m(\mathcal{D}) = d_m(\mathcal{D})^{\nicefrac{2}{|\mathcal{D}|}}$.
    We thus consider a detection criterion given by thresholds of this modified Mahalanobis distance computed over the data before and after the candidate change point w.r.t. the posterior distribution of $m_0$:
    \begin{equation}
        \Bar{d}_{m_0}(\mathcal{D}_{\tilde{c}(t):c_t}) > \nu_1 \land  \Bar{d}_{m_0}(\mathcal{D}_{c_t+1:t}) > \nu_2,
    \label{eq:cpd_criteria}
    \end{equation}
    where the thresholds $\nu_{1}, \nu_{2}$ control how strict this criterion is. A higher $\nu_i$ implies that a higher discrepancy w.r.t. the model posterior is allowed without approving a change, thus fewer change points are detected.

    This detection criterion differs from using a threshold for the distance of the complete sequence $d_{m_0}(\mathcal{D}_{\tilde{c}(t):t})$. 
	In our formulation, since we consider that the change point splits the data into two independent segments, this criterion does not include the correlation between both sequences.
    Besides, setting independent thresholds to both segments prevents false detections with outliers. For instance, if an outlier is present at the second segment $\mathcal{D}_{c_t+1:t}$, only the criterion for the second segment is satisfied, then the candidate change point is not declared as a real one.

 \subsection{Online implementation} \label{subsec:online}
   
    For each arriving batch of observations, GOCPD greedily updates the candidate change point location by optimizing the maximum likelihood problem in \eqref{eq:argmax_criteria}.     Since ternary search is a recursion, $m_1$ and $m_2$ are re-trained using parameters learned in the previous iteration as initial condition. 
    Then, GOCPD evaluates whether this data split provides enough evidence of the data not belonging to only one model (\eqref{eq:cpd_criteria}). To increase robustness to outliers, we notify a change point when \eqref{eq:cpd_criteria} is satisfied for a number  of iterations $k_{\max}$ that is context dependent. Algorithm \ref{alg:cpd} presents our proposed methodology.

    %


    
 \begin{algorithm}[t]
    \SetAlgoLined
    \SetKwInOut{Input}{Input}
    \SetKwInOut{Output}{Output}
    \textbf{Input}: $X \in \mathbb{R}^{T, D}, Y \in \mathbb{R}^{T, C}$ \\
    
     Initialize $\theta_{0,1,2}=\theta_{0,1,2}^{\text{prior}}$, $\tilde{c}(t) = 0$, $\nu_{1,2}$, $k=0$, $k_{\max}$, $t=0$, $c_t=1$, $T_{\text{ini}}$\;
     \For{$t < T$}{
     \gray{// Wait for a window $T_{\text{ini}}$ to start}\\
      \If{$t - \tilde{c}(t) < T_{\text{ini}}$}{
       \textbf{continue}\;
       }
       
        $\vx = X_{\tilde{c}(t):t}$; $\vy = Y_{\tilde{c}(t):t}$\;
     \gray{// Obtain current candidate change point}\\
       $c_t \leftarrow \textit{TernarySearch}(x, y, c_{t-1})$\;  
       
       \gray{// Evaluate change point detection criterion}\\
      \If{$d_{m_0}(\vy_{1:c_t}) > \nu_1 \land  d_{m_0}(\vy_{c_t+1:t}) > \nu_2$ \text{\textbf{and}} $c_t = c_{t-1}$}{ 
       $k \leftarrow k + 1$\;
       \If{$k > k_{\max}$}{
       Change point is detected: $\tilde{c}(t) \leftarrow c_t$\;
       Reset model parameters $\theta_{0,1,2} \leftarrow \theta_{0,1,2}^{\text{prior}}$\;
       $k \leftarrow 0\;$
       }
       }
     }
     \caption{\textbf{Greedy Online Change Point Detection (GOCPD)}}
     \label{alg:cpd}
    \end{algorithm}

    %
    %
    

\begin{figure}[t]
    \begin{minipage}[b]{\linewidth}
      \centering
      \centerline{\includegraphics[width=0.97\linewidth, trim={0 0.8cm 0.2cm 0}, clip ]{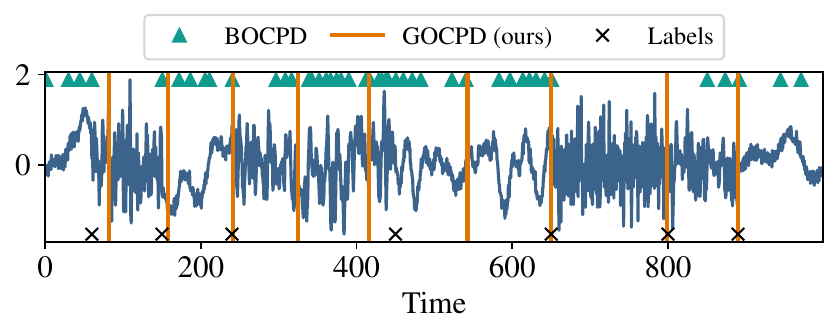}}
      \centerline{(a) Change in lengthscale.}\medskip
    \end{minipage}
    \hfill
    %
    \begin{minipage}[b]{\linewidth}
      \centering
      \centerline{\includegraphics[width=1.01\linewidth, trim={0.5cm 0.8cm 0 1.1cm}, clip ]{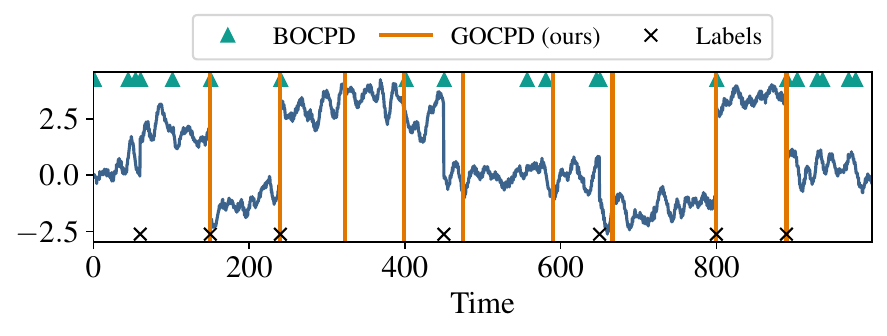} }
      \centerline{(b) Change in mean.}\medskip
    \end{minipage}
    \hfill
    \begin{minipage}[b]{\linewidth}
      \centering
      \centerline{\includegraphics[width=\linewidth, trim={0.3cm 0 0 1.1cm}, clip ]{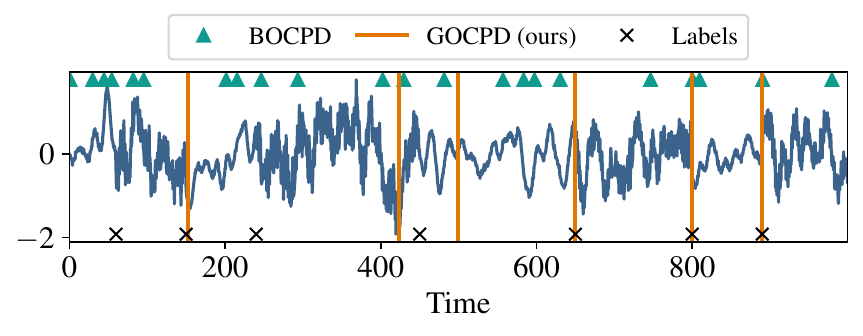}}
      \centerline{(c) Change in noise variance.}\medskip
    \end{minipage}

    \caption{\textbf{CPD on synthetic GP data.} In each figure, the change points consist of changes in each of the parameter of the RBF kernel. Overall, GOCPD returned considerably fewer false detections than BOCPD.}
    \label{fig:artif}
    \end{figure}
    
    \begin{remark}[]
    Unlike likelihood-ratio-based CPD \cite{cusum2, SPLL, cusum1}, GOCPD does not compute a likelihood \emph{ratio} but directly optimizes the likelihood of the pre- and post-change models. 
    \end{remark}

    
    \begin{remark}[]
    BOCPD \cite{adams2007bayesian} is analogous to predict the data with $m_0$ in GOCPD.
    %
    %
    Therefore, in addition to detecting changes, GOCPD also provides models capable of predicting the upcoming observation.
    After finding the change point, trained model $m_1$ can predict observations before the change point. Likewise, trained model $m_2$ can be used to forecast the data after the change point.
    \end{remark}

\section{Experimental results} \label{results}

    We evaluated GOCPD on synthetic and real-world data containing abrupt changes in the underlying dynamics. Our code is implemented on GPyTorch \cite{gpytorch}, and all experiments were executed on an Intel Core i7-9750H 2.6GHz CPU and NVIDIA GTX 1660 Ti 6Gb GPU. 
    
\subsection{Synthetic data}

    \textbf{Datasets.} We generated four time series of 1,000 points from a GP with RBF kernel and produced changes each of its hyperparameters: lengthscale $l$, noise variance $\sigma_n^2$, and mean $\mu$, at randomly chosen timestamps. 
    
    \textbf{Experimental setup.} We divided each time series with a 30\%-70\% (train-test) split, in which we used the training section to tune the hyperparameters of GOCPD. 
    We evaluated the results in terms of sensitivity, or true positive rate (TPR) and precision, or positive predictive value (PPV).

    \textbf{Results.} \figref{fig:artif} illustrates the detections made by BOCPD and GOCPD over synthetic GP time series. 
    Overall, BOCPD flagged numerous false positives, which is particularly noticeable in the time series with changes in lengthscale.
    On the other hand, for GOCPD, changes in lengthscale and mean were successfully detected.
    GOCPD also detected some false positives.
    However, although those points were not labeled change points, the dynamic of the underlying GP at those points clearly changed, e.g., at $t=400$ in the time series with changes in mean and at $t\approx550$ in the time series with changes in lengthscale. 
    \figref{fig:artif}c shows the detections for the case of change in the noise magnitude, where GOCPD did not detect the first and third change points. However, this corroborates the robustness of GOCPD as it was capable of effectively recognizing the change in variance as simply noise.
    
    \tabref{tab:rates} shows the results on synthetic datasets. 
    BOCPD achieves perfect TPR in two datasets, but its PPV is considerably lower.
    In contrast, our method returns a lower TPR than that of BOCPD, but consistently outperforms BOCPD in terms of PPV for all synthetic time series, i.e., most of the detected changes of GOCPD correspond to true changes.

    \begin{table}[t]
    \centering
    \footnotesize
    \resizebox{\linewidth}{!}{\begin{tabular}{l cc cc cc} 
         \toprule
         \multirow{2}{*}{\textbf{Method}} & \multicolumn{2}{c}{Change in $\ell$} & \multicolumn{2}{c}{Change in $\mu$} & \multicolumn{2}{c}{Change in $\sigma_n^2$} \\ 
         \cmidrule{2-3} \cmidrule{4-5} \cmidrule{6-7} 
         & TPR & PPV & TPR & PPV & TPR & PPV\\
         \midrule
         BOCPD & $0.86$ & $0.13$ & $\mathbf{1.00}$ & $0.33$ & $\mathbf{0.86}$ & $0.26$\\
         GOCPD (ours) & $\mathbf{1.00}$ & $\mathbf{0.78}$ & $0.86$ & $\mathbf{0.67}$ & $0.71$ & $\mathbf{0.83}$ \\
         \bottomrule
    \end{tabular}}
    \caption{\textbf{True positive rate (TPR) and positive predictive value (PPV) in synthetic data.} GOCPD outperforms BOCPD in terms of PPV overall.} \label{tab:rates}
    \end{table}

\subsection{Real world data}

\begin{figure}[t]
    \centering
    \includegraphics[width=0.99\linewidth]{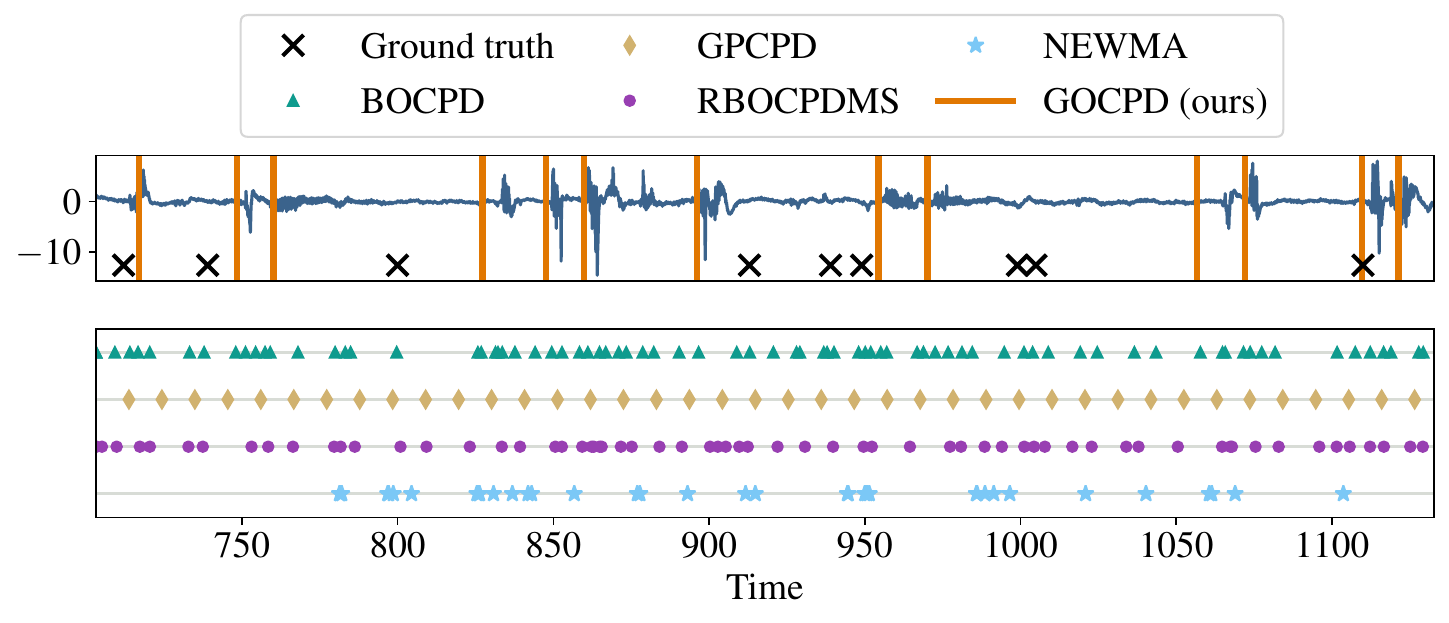}
    \caption{\textbf{Change detection on EEG data}. GOCPD returns the lowers false discovery rate.}
    \label{fig:eeg}
\end{figure}

\begin{figure}[t]
    \centering
    \includegraphics[width=0.95\linewidth]{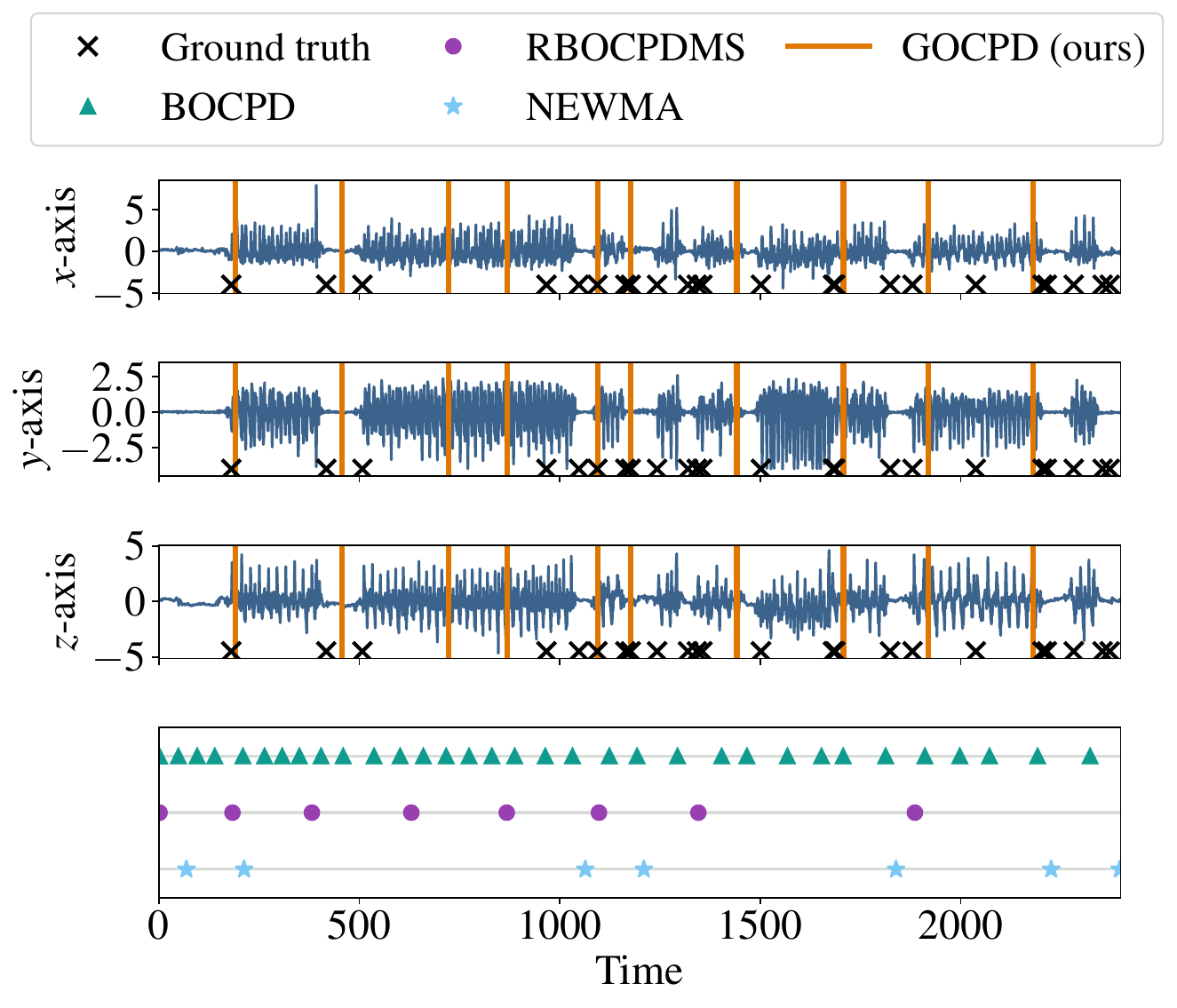}
    \caption{\textbf{Human activity data}. GOCPD achieves the lowest false detection rate, while keeping a desired true detection rate.}
    \label{fig:multi_act}
\end{figure}

    \textbf{Datasets.} GOCPD was also applied to three real-world settings: the Well-log dataset \cite{well_log} is a classic benchmark of univariate CPD that records 4,050 nuclear magnetic resonance levels while drilling a well; the Neonatal electroencephalography (EEG) dataset \cite{eeg_dataset} contains multivariate EEG signals with brain seizures; and the Human activity dataset\footnote{\url{http://hasc.jp/hc2011/index-en.html}} (HAct) provides multivariate human movement signals collected by three-axis accelerometers. For Human activity, we evaluated the CPD methods in both univariate and multivariate scenario.
    
    \textbf{Baselines.} For GOCPD, we use Gaussian processes \cite{rasmussen} for the univariate datasets and Multi-output Gaussian processes \cite{mogp} for the multivariate dataset.
    We consider RBF kernels for all datasets except for Well-log, in which we use the Dirac delta function for the covariance $K(x, x') = \mathbbm{1}_{x=x'}$ due to observed uncorrelatedness.
    
    We compared GOCPD against three Bayesian methods (BOCPD \cite{adams2007bayesian}, GPCPD \cite{saaturras2010}, and RBOCPDMS \cite{knoblauch2018doubly})  and the model-free approach NEWMA \cite{newma}. 
    %
    %
    For BOCPD, RBOCPD, and NEWMA, we relied on officially released codes.
    
    \textbf{Experimental setup.} 
    %
    To quantitatively compare the number of false detections, we report the false discovery rate $\text{FDR} = 1 - \text{PPV}$, which measures the proportion of false detections in the total set of detected changes.
    For all methods, we report the averaged values in five independent runs, in which the parameters of the models are randomly initialized. 
    To visualize the detected changes of each method, we select the best performing run among the five repetitions.


\begin{table}[t]
    \centering
    \footnotesize
    \resizebox{\linewidth}{!}{
    \begin{tabular}{l cc cc} 
         \toprule
         \textbf{Method} & \textbf{EEG} & \textbf{HAct.} (uni) & \textbf{HAct.} (multi) & \textbf{Well-log} \\ 
         \midrule
         BOCPD & $0.88$ & $0.33$ & $0.47$ & $0.53$ \\
         RBOCPDMS & $0.86$ & $0.22$ & $0.33$ & $\mathbf{0.30}$ \\
         NEWMA & $0.81$ & $0.25$ & $0.14$ & $0.71$ \\ \midrule
         GOCPD (ours) & $\mathbf{0.54}$ & $\mathbf{0.15}$ & $\mathbf{0.10}$ & $\mathbf{0.30}$ \\
         \bottomrule
    \end{tabular}}
    \caption{\textbf{False discovery rates (FDR) in real-world datasets.} GOCPD achieved the lowest FDR in all datasets.} \label{tab:real_rates}
    \end{table}

   \begin{table}[t]
    \centering
    \footnotesize
    \resizebox{\linewidth}{!}{
    \begin{tabular}{l cc} 
         \toprule
         \textbf{Method} & \textbf{Time complexity} & \textbf{Time/\#points} [s]\\ 
         \midrule
         BOCPD & $\mathcal{O}(T)$ & $\underline{0.156}$ \\
         GPCPD & $\mathcal{O}(T^4)$ & $0.404$ \\
         RBOCPDMS  & $\mathcal{O}(1)$ & $13.293$ \\
         \midrule
         GOCPD (ours) + GP & $\mathcal{O}(T^3 \log(T))$ & $\mathbf{0.075}$ \\
         \bottomrule
    \end{tabular} }
    \caption{\textbf{Average CPU execution time on Well-log dataset}. 
    Approximately, GOCPD performed $177$ times faster than RBOCPDMS, $5$ times faster than GPCPD, and $2$ times faster than BOCPD.} \label{tab:time}
    \end{table} 
    
    \textbf{Results.} \figref{fig:eeg} shows the detected changes for the EEG dataset.
    Most of the detections of NEWMA corresponded to false positives, and also false negatives before $t=770$.
    In turn, BOCPD and RBOCPDMS detected all changes in the ground truth. 
    However, aside from the ground truth, most detections were false positives, even when the signal did not have any evident change, e.g., for $t\in(1000, 1050)$.
    GPCPD also mostly returned false discoveries. 
    %
    %
    %
    In contrast, GOCPD flagged considerably fewer detections.
    The false negatives of GOCPD generally corresponded to transitions between high-variance data to low-variance data, e.g., the changes at $t\approx800$ and $t\approx1000$.
    %
    %
    %
    Recall that the annotations only partition the signal into seizure and non-seizure segments.
    However, noticeable changes may also arise inside both categories, for instance, in the intervals $(800, 900)$ and the remaining interval after $t=1100$, which were detected by GOCPD.
    Notably, our method did not notify a change whenever an abrupt jump appeared, e.g., for $t \in (860, 900)$.
    These results highlight that our proposal is more robust than model-free and BOCPD-based methods.
    
    \figref{fig:multi_act} illustrates the result for the multivariate Human Activity dataset. 
    Here, most of the results of NEWMA were real changes, but it also missed several labeled changes.
    In this case, most of the detections of BOCPD corresponded to false discoveries, as we discussed in \subsecref{subsec:previous_Work}, while GOCPD only flagged two false detections.
    Also, 
    %
    RBOCPDMS produced false negatives possibly due to the use of $\beta$-divergences. This was because, dending on $\beta$, RBOCPDMS may  consider extreme abrupt changes as outliers, thus leading to false negatives.
    
%
%
    \tabref{tab:real_rates} shows the detection results for the real-world datasets that provide labels of the true change points.
    Importantly, GOCPD achieved the lowest FDR for all datasets.
    In particular, for the multivariate Human Activity dataset, the FDR of GOCPD was less than $25\%$ of that of BOCPD.
    In contrast, BOCPD flagged the highest FDR, followed by RBOCPDMS. 
    %
    %

\section{Discussion} \label{sec:discussion}

\subsection{Time complexity} \label{subsec:time}

    For a fair comparison, we executed GOCPD on a CPU and used the original version of BOCPD and RBOCPDMS without pruning.
    We kept model-free methods out of our analysis since they need not to model the observations, thus their execution times are expected to be shorter.
    
    For each CPD method, \tabref{tab:time} shows the theoretical complexity and the running times for the Well-log experiment. Though GOCPD is theoretically expected to have a larger  complexity than BOCPD and RBOCPDMS, this was not the case in practice. This can be attributed to our implementation where GOCPD does not search over the complete segment of the data, leveraging \propref{prop:save_cand}. Therefore, the theoretical complexity of GOCPD accounts for the worst-case scenario, in which the saved candidate change point of GOCPD always stays at $t=0$, which is improbable in practice. We present a deeper discussion of this result in the next subsection.

    \tabref{tab:time} shows that GOCPD is the fastest method overall, followed by BOCPD. 
    However, since we implement GOCPD with Gaussian processes, a fair comparison is against GPCPD, where GOCPD is at least $5\times$ faster.
    Nonetheless, even comparing our method against RBOCPDMS, which uses GLR as the underlying predictive model, GOCPD presents a $177\times$ speedup.
    %
    These figures highlight the advantage of minimizing the searching interval by storing the best candidate at each iteration and using ternary search to identify the optimal location of the change point. 

\subsection{Effective number of evaluated timestamps} \label{subsec:evaluations}
    
\begin{figure}[t]
        \centering
        \includegraphics[width=\linewidth]{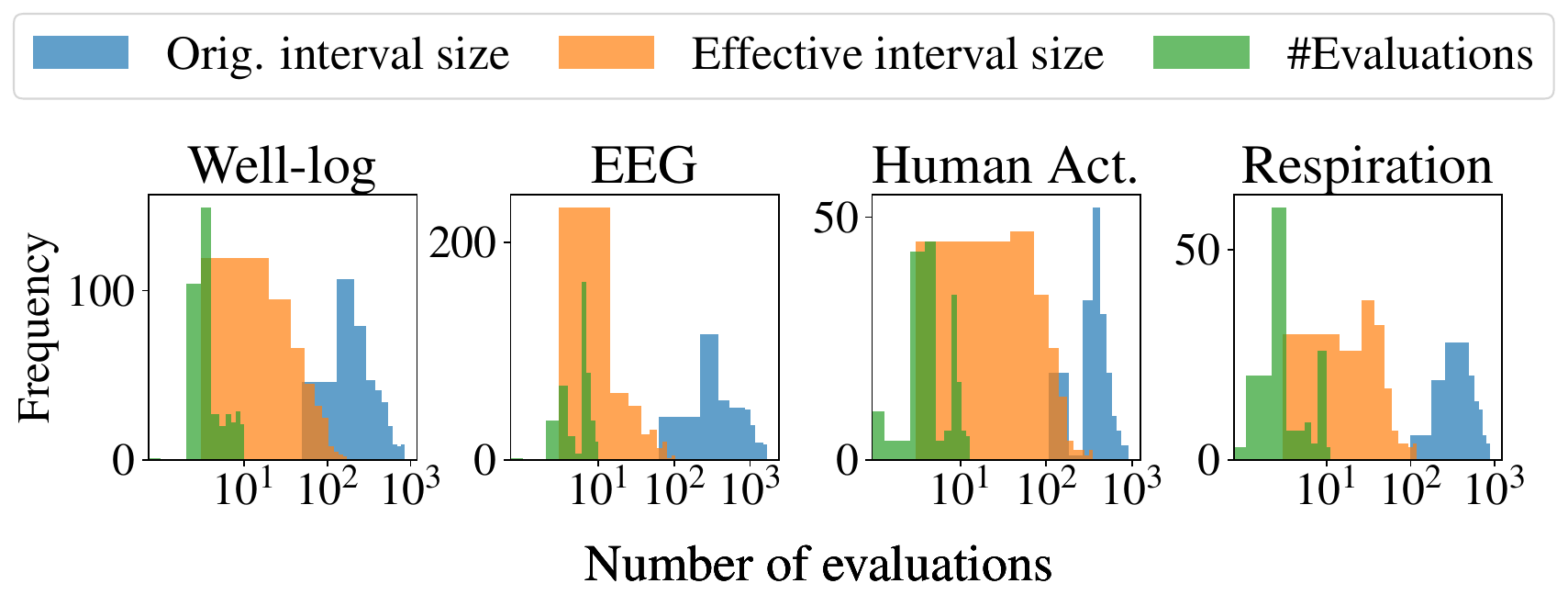}
        \caption{\textbf{Histogram of original interval size, effective interval size, and real number of evaluations.} Notably, by saving the candidate and using ternary search to maximize \eqref{eq:argmax_criteria}, the number of evaluations decreased in two orders of magnitude.}
        \label{fig:interval_comparison}
    \end{figure}

    For each dataset, \figref{fig:interval_comparison} shows the distribution of the original interval size at each iteration, i.e., $t-\tilde{c}(t)$, and the effective interval size after saving the candidate change point, i.e., $t-c_t$. 
    %
	%
	In practice, we observe that $t-c_t \ll t-\tilde{c}(t)$ in the datasets studied in this work.
    The difference of both distributions is of at least one order of magnitude, which illustrates the benefit of saving the candidate change point at each iteration.

    Critically, the difference between the distribution of the effective interval size and that of the real number of evaluations per iteration demonstrates the effect of optimizing the change point location metric using ternary search.
    Overall, joining both components, GOCPD decreases the number of evaluations needed to optimize the change point location metric in approximately two orders of magnitude with respect to the original case, that has linear time complexity for each batch.
    
    
\subsection{Limitations} \label{subsec:limitations}

    GOCPD is motivated by the intuition of searching a candidate change point in the thus-far observed data before the current time and waiting for a minimum period to effectively notifiy a change point. This procedure increases the robustness of detection, but also creates an irreducible detection delay. There is a trade-off between robustness and detection delay for tuning this hyperparameter. 
    In addition, experimental results show that, using GPs, changes from high to small variance are more difficult to detect with GOCPD, since a model with high variance ($m_1$) can still fit data with lower variance, thus the likelihood of $m_1$ does not decrease abruptly.
    %

\section{Conclusion} \label{conclusion}

    In this paper, we have introduced GOCPD, an online CPD method that uses two simple criteria simulating human intuition when making online change detection. 
    GOCPD uses a computationally efficient searching mechanism that leverages the unimodal property of the probability of the observations coming from two independent distributions. By greedily saving candidate change points, the space of possible change points is considerably reduced.
    We have shown that our method can effectively detect changes in synthetic datasets and outperforms existing methods in real-world data in terms of false discovery rates. 
    %
    %
    Since our method recycles trained models needed for the searching algorithm, future work includes online models that only need to be retrained over the new set of observations \cite{streaming}.
    Additionally, at each step of the ternary search, one model \textit{looks forward} and the other model \textit{looks backward} and thus needs to \textit{forget} datapoints. Therefore, a deeper analysis into models that better forget data are a promising future research direction.

\bibliographystyle{IEEEtran}
\bibliography{IEEEabrv,refs}

\end{document}